\documentclass[conference]{IEEEtran}
\ifCLASSINFOpdf
\else
\fi
\usepackage[cmex10]{amsmath}
\usepackage{url}

\usepackage[]{fancyhdr} %
\newcommand{\changefont}{\fontsize{9}{9}\selectfont}
\usepackage{amsmath}
\usepackage{booktabs}
\usepackage{cite}
 \usepackage[pdftex]{graphicx}
 \usepackage{tikz}
\usetikzlibrary{shapes.geometric}

\pgfdeclareplotmark{mystar}{
    \node[star,star point ratio=2.25,minimum size=5pt,
          inner sep=0pt,draw=blue,solid,fill=blue] {};
}
\usetikzlibrary{plotmarks}
\usetikzlibrary{calc}
\usetikzlibrary{arrows}
\usetikzlibrary{arrows,intersections}
\usetikzlibrary{arrows.meta}
\usetikzlibrary{shapes,decorations}
\usetikzlibrary{positioning}
\usepackage{pgfplots}
\tikzset{sin v source/.style={
  circle,
  draw,
  append after command={
    \pgfextra{
    \draw
      ($(\tikzlastnode.center)!0.5!(\tikzlastnode.west)$)
       arc[start angle=180,end angle=0,radius=0.425ex]
      (\tikzlastnode.center)
       arc[start angle=180,end angle=360,radius=0.425ex]
      ($(\tikzlastnode.center)!0.5!(\tikzlastnode.east)$)
    ;
    }
  },
  scale=1.5,
 }
}
\usetikzlibrary{pgfplots.groupplots, backgrounds, plotmarks, calc, spy, pgfplots.polar, shapes.geometric, arrows, external, patterns}

\tikzset{
    invisible/.style={opacity=0},
    visible on/.style={alt={#1{}{invisible}}},
    alt/.code args={<#1>#2#3}{%
      \alt<#1>{\pgfkeysalso{#2}}{\pgfkeysalso{#3}} 
    },
  }

\tikzset{sin v source/.style={
  circle,
  draw,
  append after command={
    \pgfextra{
    \draw
      ($(\tikzlastnode.center)!0.5!(\tikzlastnode.west)$)
       arc[start angle=180,end angle=0,radius=0.425ex] 
      (\tikzlastnode.center)
       arc[start angle=180,end angle=360,radius=0.425ex]
      ($(\tikzlastnode.center)!0.5!(\tikzlastnode.east)$) 
    ;
    }
  },
  scale=1.5,
 }
}

\pgfplotsset{compat=1.14}

 \usepackage{amstext,epsfig,amssymb,amsbsy,amsmath}
\usepackage{cases}
\newcommand{\mb}{\mathbf}
\usepackage{bm}
\IEEEoverridecommandlockouts
\begin{document}

%
\title{Neural Networks for Encoding \\ Dynamic Security-Constrained Optimal Power Flow }

\author{
\IEEEauthorblockN{Ilgiz~Murzakhanov, Andreas~Venzke, George~S.~Misyris, Spyros~Chatzivasileiadis}
\IEEEauthorblockA{Department of Wind and Energy Systems \\
Technical University of Denmark\\
Kgs. Lyngby, Denmark\\
\{ilgmu, andven, gmisy, spchatz\}@dtu.dk}
\thanks{This work is supported by the ID-EDGe project, funded by Innovation Fund Denmark, Grant Agreement No. 8127-00017B, by the FLEXGRID project, funded by the European Commission Horizon 2020 program, Grant Agreement No. 863876, and by the ERC Starting Grant VeriPhIED, Grant Agreement No. 949899.}}


\maketitle
\thispagestyle{fancy}
\pagestyle{fancy}

\begin{abstract}
This paper introduces a framework to capture previously intractable optimization constraints and transform them to a mixed-integer linear program, through the use of neural networks. We encode the feasible space of optimization problems characterized by both tractable and intractable constraints, e.g. differential equations, to a neural network. Leveraging an exact mixed-integer reformulation of neural networks, we solve mixed-integer linear programs that accurately approximate solutions to the originally intractable non-linear optimization problem. We apply our methods to the AC optimal power flow problem (AC-OPF), where directly including dynamic security constraints renders the AC-OPF intractable. Our proposed approach has the potential to be significantly more scalable than traditional approaches. We demonstrate our approach for power system operation considering N-1 security and small-signal stability, showing how it can efficiently obtain cost-optimal solutions which at the same time satisfy both static and dynamic security constraints.
\end{abstract}


\IEEEpeerreviewmaketitle

\section{Introduction}
\subsection{Motivation}
In a wide range of optimization problems, especially related to physical systems, the feasible space is characterized by differential equations and other intractable constraints \cite{agrawal2013optimization}. Inspired by power system operation, this paper uses the AC optimal power flow (AC-OPF) problem with dynamic security constraints as a guiding example to introduce a framework that efficiently captures previously intractable constraints and transforms them to a mixed-integer linear program, through the use of neural networks. More specifically, we use neural networks to encode the previously intractable feasible space, and through an exact transformation we convert them to a set of linear constraints with binary variables that can be integrated to an equivalent mixed-integer linear problem (MILP).


Power system security assessment is among the most fundamental functions of power system operators. With growing uncertainty both in generation and demand, the complexity of this task further increases, necessitating the development of new approaches \cite{panciatici2012operating}. In power systems, the AC optimal power flow (AC-OPF) is an essential tool for cost-optimal and secure power system operation  \cite{cain2012history}. The non-convex AC-OPF problem minimizes an objective function (e.g. generation cost) subject to steady-state operational constraints (e.g. voltage magnitudes and transmission line flows). While the obtained generation dispatch from the AC-OPF solution ensures compliance with static security criteria such as N-1 security, the dispatch additionally needs to comply with dynamic security criteria such as small-signal or transient stability. However, directly including dynamic security constraints renders the AC-OPF problem intractable \cite{capitanescu2011state}. To obtain solutions which satisfy both static and dynamic security criteria, we propose a novel framework using neural networks to encode dynamic security constrained AC-OPF to MILPs.

\subsection{Literature Review}
In literature, a range of works \cite{zarate2009securing, xu2012hybrid, vaahedi2001dynamic,condren2006expected} have proposed iterative approaches and approximations to account for dynamic security constraints in AC-OPF problems. For a comprehensive review please refer to \cite{capitanescu2011state}. The work in \cite{zarate2009securing} considers transient stability by discretizing a simplified formulation of the power system dynamics and proposes an iterative solution scheme. Alternatively, to include transient stability constraints, the work in \cite{xu2012hybrid} proposes a hybrid solution approach using evolutionary algorithms. The work in \cite{vaahedi2001dynamic} addresses voltage stability and proposes a three-level hierarchical scheme to identify suitable preventive and corrective control actions. To approximate the small-signal stability criterion, the work in \cite{condren2006expected} linearizes the system state around a given operating point and includes the eigenvalue sensitivities in the AC-OPF problem. 
While the majority of these approaches are tailored to a specific dynamic stability criterion and require to iteratively solve non-linear programs (NLPs), in this paper we propose a general framework which allows us to encode any security criterion. 


A range of machine learning approaches have been proposed to learn optimal solutions to the AC-OPF problem \cite{Dalal2017,fioretto2019predicting} and approximations thereof such as the DC-OPF problem \cite{pan2019deepscopf,chen2020learning}, without considering dynamic security constraints.  
The work in \cite{Dalal2017} compares different machine learning approaches and finds that neural networks achieve the best performance. To predict optimal solutions to AC-OPF problems, the work in \cite{fioretto2019predicting} trains deep neural networks and penalizes constraints violations. For the DC-OPF approximation, the work in \cite{pan2019deepscopf} trains deep neural networks as well, achieving a speed-up of two orders of magnitude compared to conventional methods and accounting for static security constraints. Instead of directly learning the optimal solutions, the work in \cite{chen2020learning} predicts the set of active constraints for the DC-OPF approximation. Note that the DC-OPF approximation neglects reactive power, voltage magnitudes and losses and can lead to substantial errors \cite{dvijotham2016error}. Ref. \cite{9281941} is among the few that employ a multi-input multi-output random forest model to solve a security-constrained AC-OPF. The model first solves voltage magnitudes and angles and then calculates other parameters, such as power generation settings; still, this work does not include any dynamic security constraints. While these approaches demonstrate substantial computational speed-up compared to conventional solvers, they do not account for dynamic security criteria, and most of them rely on a large training dataset of computed optimal solutions to the AC-OPF problem. Obtaining such a dataset is computationally prohibitive for dynamic security-constrained AC-OPF.

Using machine learning techniques for static and dynamic security assessment has been explored  in literature, for a comprehensive survey please refer to \cite{wehenkel2012automatic}.  The main area of application has been the screening of a large number of operating points with respect to different security criteria. Several works have shown that the proposed machine learning methods including neural networks require only a fraction of the computational time needed for conventional methods. Fewer works \cite{cremer2018data,halilbavsic2018data,gutierrez2010neural} explored using machine learning techniques to directly include dynamic security constraints in AC-OPF problems. The work in \cite{cremer2018data} proposed ensemble decision trees to identify corrective actions to satisfy security criteria. Embedding the decision trees in an AC-OPF framework requires to solve computationally highly expensive mixed-integer non-linear problems (MINLPs). To address computational tractability, the work in \cite{halilbavsic2018data} proposed a second-order cone relaxation of the AC-OPF to relax the MINLP to a mixed-integer second-order cone program (MISOCP). This, however, does not guarantee feasibility of the obtained solution in case the relaxation is inexact. Approaching mixed-integer programming (MIP) problem with the use of a convolutional neural network (CNN) is presented in \cite{9524441}. The proposed solution targets scenario-based security-constrained unit commitment (SCUC) with a battery energy storage system (BESS) and has two stages. In the first stage, CNN-SCUC trains a CNN to obtain solutions to the binary variables corresponding to unit commitment decisions. In the second stage, the continuous variables corresponding to unit power outputs are solved by a small-scale convex optimization problem. The proposed model-free CNN-SCUC algorithm greatly reduces the computational complexity and obtains close-to-optimal solutions. The work in \cite{gutierrez2010neural} uses a non-linear representation of neural networks with one single hidden layer to represent the stability boundary in the AC-OPF problem. Instead of solving computationally expensive NLPs or MINLPs, we propose a novel framework using neural networks to encode the dynamic security constrained AC-OPF to MILPs.

\subsection{Main Contributions}
 The main contributions of our work are:  

\begin{enumerate}
        \item[1)] Using classification neural networks we encode the feasible space of AC-OPF problems including any type of static and dynamic security criteria.
        \item[2)] Leveraging a mixed-integer linear reformulation of the trained neural network and a systematic iterative procedure to include non-linear equality constraints, we accurately approximate cost-optimal solutions to the original intractable AC-OPF problems. 
        \item[3)] We introduce a method to trade-off conservativeness of the neural network prediction with cost-optimality, ensuring feasibility of the obtained solutions.
    \item[4)] Considering both N-1 security and small-signal stability, and using an IEEE 14 bus system, we demonstrate how the proposed approach allows to obtain cost-optimal solutions which at the same time satisfy both static and dynamic security constraints.
\end{enumerate}
\subsection{Outline}
The structure of this paper is as follows: In Section~II, we formulate optimization problems with intractable dynamic security constraints including the dynamic security-constrained AC optimal power flow. In Section~III, we encode the feasible space using neural networks, state the mixed-integer linear reformulation of the trained neural network, and introduce efficient methods to include equality constraints and to ensure feasibility. In Section~IV, we demonstrate our methods using an IEEE 14 bus system and considering both N-1 security and small-signal stability. Section~V concludes.

\section{Non-linear Optimization Problems with Dynamic Security Constraints}
\subsection{General Formulation} We consider the following class of non-linear optimization problems with intractable dynamic constraints:
\begin{alignat}{2}
    \min_{\mb{x} \in \mathcal{X},\mb{u} \in \mathcal{U}} \quad & f(\mb{u}) &&  \label{obj1} \\
    \text{s.t.}  \quad & g_i (\mb{x},\mb{u}) \leq 0  && \quad \forall \, i = 1, ...,m \label{ineq} \\  
    & h_i (\mb{x},\mb{u}) = 0  && \quad \forall \, i = 1, ...,n \label{eq} \\
    & \phi_i(\mb{x},\mb{u}) \in \mathcal{S}_i && \quad \forall \, i = 1, ...,l \label{intract} 
\end{alignat}
The variables are split into state variables $\mb{x}$ and control variables $\mb{u}$ which are constrained to the sets $\mathcal{X}$ and $\mathcal{U}$, respectively. The objective function is denoted with $f$ in \eqref{obj1}. There is a number of $m$ and $n$ non-linear inequality and equality constraints $g$ and $h$, respectively. The $l$ number of constraints $\phi$ in \eqref{intract} encode dynamic security constraints (e.g. based on differential equations) and are intractable. While the non-linear optimization problem \eqref{obj1}--\eqref{eq} can be solved using a non-linear solver, the addition of the constraints in \eqref{intract} renders the optimization problem intractable. Note that for given fixed system state $(\mb{x},\mb{u})$ we can determine computationally efficiently whether it belongs to set $\mathcal{S}$ satisfying dynamic constraints, i.e. is feasible or infeasible. One possible solution strategy is to follow an iterative approach where the optimization problem \eqref{obj1}--\eqref{eq} is solved first, and then if constraint \eqref{intract} is violated a nearby solution is identified which satisfies \eqref{intract}. This has several drawbacks including that this procedure does not guarantee the recovery of a solution that is feasible to the original problem, and the feasibility recovery procedure does not consider optimality of the solution. In the following, using power system operation as a guiding example, we will present an approach that allows to directly approximate a high-quality solution to \eqref{obj1}--\eqref{intract} by solving mixed-integer linear programs instead.
\subsection{Application to AC Optimal Power Flow (AC-OPF)}
In the following, we consider the AC-OPF problem with combined N-1 security and small-signal stability criteria as guiding example. Note that our proposed methodology is general and can include any static or dynamic security criteria. The AC-OPF problem optimizes the operation of a power grid consisting of a set of $\mathcal{N}$ buses connected by a set of $\mathcal{L}$ lines. A subset of the set of $\mathcal{N}$ buses has a generator connected and is denoted with $\mathcal{G}$. The variables are the vectors of active and reactive power injections $\mb{p}$ and $\mb{q}$, the voltage magnitudes $\mb{v}$ and voltage angles $\theta$. Each of these vectors have the size $n\times1$, where $n$ is the number of buses in the set $\mathcal{N}$. The control variables $\mb{u}$ are the active power injections and voltage set-points of generators $\mb{p_g}$ and $\mb{v_g}$, i.e. the entries in $\mb{p}$ and $\mb{v}$  that correspond to the buses in the set $\mathcal{G}$. If the control variables are fixed, the remaining state variables can be identified by solving an AC power flow, i.e. solving a system of non-linear equations \cite{Tinney67}. To satisfy the N-1 security criterion, the identified control variables  $\mb{u}$ must lead to a power system state which complies with the operational constraints for a set of line outages. We denote this set with $\mathcal{C}$, with the first element $\mathcal{C}_1 = \{0\}$ corresponding to the intact system state. The dimension of set $\mathcal{C}$ is defined by the number of considered contingencies plus the intact system state. As for its structure, the set $\mathcal{C}$ contains the IDs of the outaged lines, while for the intact system state the corresponding ID is '0'. The superscript '$c$' in \eqref{con_p}--\eqref{SS2} denotes the corresponding outaged system state and the superscript '$0$' the intact system state. To ensure small-signal stability, we analyze the stability of the linearization of the power system dynamics around the current operating point \cite{milano2010power}. The system matrix $\mb{A}$ describes the linearized system dynamics as function of the current operating point defined by $(\mb{p},\mb{q},\mb{v},\bm{\theta})$. To satisfy the small-signal stability criterion, the minimum damping ratio associated with the eigenvalues $\bm{\lambda}$ of $\mb{A}$ has to be larger than $\gamma_{\text{min}}$; where $\gamma_{\text{min}} \leq \min_{\bm{\lambda}} \tfrac{-\Re\{\bm{\lambda}\}}{\sqrt{\Im\{\bm{\lambda}\}^2 + \Re\{\bm{\lambda}\}^2}}$.  Finally, we formulate the N-1 security and small-signal stability constrained AC-OPF:
\begin{alignat}{2}
\min_{\mb{p}^c,\mb{q}^c,\mb{v}^c,\bm{\theta}^c,\bm{\lambda}^c,\bm{\nu}^c} \, \, & f(\mb{p_g^0}) && \label{obj}\\
    \text{s.t.}  \quad& \mb{p}^{\text{min}} \leq \mb{p}^c \leq \mb{p}^{\text{max}} && \, \, \forall c \in \mathcal{C} \label{con_p} \\
    & \mb{q}^{\text{min}} \leq \mb{q}^c \leq \mb{q}^{\text{max}} && \, \,\forall c \in \mathcal{C} \label{con_q} \\
    & \mb{v}^{\text{min}} \leq \mb{v}^c \leq \mb{v}^{\text{max}} && \,  \, \forall c \in \mathcal{C} \label{con_v} \\
    & | \mb{s}_{\text{line}} (\mb{p}^c,\mb{q}^c,\mb{v}^c,\bm{\theta}^c) | \leq \mb{s}_{\text{line}}^{\text{max}} && \, \, \forall c \in \mathcal{C} \label{con_s} \\
    & \mb{s}_{\text{balance}} (\mb{p}^c,\mb{q}^c,\mb{v}^c,\bm{\theta}^c) = \mb{0} && \, \, \forall c \in \mathcal{C} \label{con_acpf} \\
     & \mb{p_g^0} = \mb{p_g^c}, \, \mb{v_g^0} = \mb{v_g^c} && \, \, \forall c \in \mathcal{C} \label{link} \\
      & \mb{A}(\mb{p}^c,\mb{q}^c,\mb{v}^c,\bm{\theta}^c) \bm{\nu}^c = \bm{\lambda}^c \bm{\nu}^c && \, \, \forall c \in \mathcal{C}  \label{SS1} \\
      & \gamma_{\text{min}} \leq \min_{\bm{\lambda}^c} \tfrac{-\Re\{\bm{\lambda}^c\}}{\sqrt{\Im\{\bm{\lambda}^c\}^2 + \Re\{\bm{\lambda}^c\}^2}}&&\, \, \forall c \in \mathcal{C}  \label{SS2} 
\end{alignat}
The objective function in \eqref{obj} minimizes the cost of the system generation for the intact system state. All inequality and equality constraints \eqref{con_p}--\eqref{SS2} have to hold for the intact system state and each contingency in $\mathcal{C}$. The inequality constraints in \eqref{con_p}--\eqref{con_v} define minimum and maximum limits on active and reactive power injections and voltage magnitudes. The inequality constraint in \eqref{con_s} bounds the absolute apparent branch flow $\mb{s}_{\text{line}}$ for each line in $\mathcal{L}$. The equality constraint in \eqref{con_acpf} enforces the AC power flow balance $\mb{s}_{\text{balance}}$ at each bus $\mathcal{N}$ in the system. For a detailed mathematical description of the apparent branch flow and the AC power flow balance, for brevity, please refer to \cite{cain2012history}. We consider preventive control for N-1 security defined in \eqref{link}, i.e. the generator active power set-points and voltage set-points remain fixed if a line outage occurs. The inequality constraint \eqref{SS1} ensures a minimum damping ratio of critical modes in the system. The term $\bm{\nu}$ denotes the right hand eigenvectors of the system matrix $\mb{A}$. Note that the small-signal stability constraints \eqref{SS1}--\eqref{SS2} have to hold for each considered line outage in set $\mathcal{C}$. For more details related to small-signal stability, we refer the interested reader to \cite{8008602}. Adding the small-signal stability constraints in \eqref{SS1}--\eqref{SS2} increases the non-linearity and computational complexity of the optimization problem substantially, rendering it intractable and requiring iterative solution approaches \cite{condren2006expected}. In the following, we introduce a framework which allows to directly approximate cost-optimal solutions to \eqref{obj}--\eqref{SS2}. 

\section{Methodology to Accurately Approximate Cost-Optimal and Feasible Solutions}

\subsection{Encoding Feasible Space using Neural Networks}
We denote the feasible space $\mathcal{F}$ of the optimization problem \eqref{obj1}--\eqref{intract} as: $\mathcal{F}:=\{\mb{x} \in \mathcal{X} \text{ and } \mb{u} \in \mathcal{U} \text{ satisfy \eqref{ineq}--\eqref{intract}} \}$. As first step, we train a classification neural network which takes the control variables $\mb{u} \in \mathcal{U}$ as input and predicts whether there exists state variables $\mb{x} \in \mathcal{X}$ such that the resulting operating point is in the feasible space $(\mb{x},\mb{u}) \in \mathcal{F}$ or is infeasible $(\mb{x},\mb{u}) \notin \mathcal{F}$.
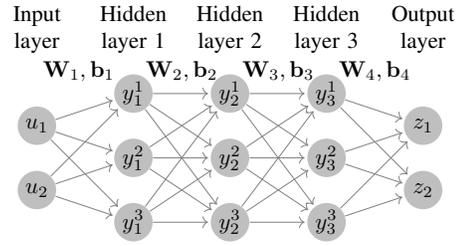
\begin{figure}
  \def\layersep{1.5cm}
\centering
\resizebox{0.75\columnwidth}{!}{%
\begin{tikzpicture}[shorten >=1pt,->,draw=black!50, node distance=\layersep]
    \tikzstyle{every pin edge}=[<-,shorten <=1pt]
    \tikzstyle{neuron}=[circle,fill=black!25,minimum size=17pt,inner sep=0pt]
    \tikzstyle{input neuron}=[neuron];
    \tikzstyle{output neuron}=[neuron];
    \tikzstyle{hidden neuron}=[neuron];
    \tikzstyle{annot} = [text width=4em, text centered]

    \foreach \name / \y in {1,...,2}
        \node[input neuron] (I-\name) at (0,-\y) {$u_\y$};

    \foreach \name / \y in {1,...,3}
        \path[yshift=0.5cm]
            node[hidden neuron] (H1-\name) at (\layersep,-\y cm) {$y^\y_1$};
            
    \foreach \name / \y in {1,...,3}
        \path[yshift=0.5cm]
            node[hidden neuron,right of=H1-\name] (H2-\name) at (\layersep,-\y cm) {$y^\y_2$};
            
    \foreach \name / \y in {1,...,3}
        \path[yshift=0.5cm]
            node[hidden neuron,right of=H2-\name] (H3-\name) at (\layersep+\layersep,-\y cm) {$y^\y_3$};

    \node[output neuron,right of=H3-2,yshift=0.5cm] (O1) {$z_1$};
    \node[output neuron,right of=H3-3,yshift=0.5cm] (O2) {$z_2$};
    \foreach \source in {1,...,2}
        \foreach \dest in {1,...,3}
            \path (I-\source) edge (H1-\dest);
            
                \foreach \source in {1,...,3}
        \foreach \dest in {1,...,3}
            \path (H1-\source) edge (H2-\dest);
            
                \foreach \source in {1,...,3}
        \foreach \dest in {1,...,3}
            \path (H2-\source) edge (H3-\dest);

    \foreach \source in {1,...,3}
        \path (H3-\source) edge (O1);
        
    \foreach \source in {1,...,3}
        \path (H3-\source) edge (O2);

    \node[annot,above of=H1-1, node distance=1cm] (hl1) {Hidden layer 1};
    \node[annot,above of=H1-1, node distance=0.35cm] (w1b1) {};
    \node[annot,left of=w1b1, node distance=0.85cm] (w1b1t) {$\mb{W}_1,\mb{b}_1$};
    \node[annot,above of=H2-1, node distance=1cm] (hl2) {Hidden layer 2};
        \node[annot,above of=H2-1, node distance=0.35cm] (w2b2) {};
    \node[annot,left of=w2b2, node distance=0.75cm] (w2b2t) {$\mb{W}_2,\mb{b}_2$};
    \node[annot,above of=H3-1, node distance=1cm] (hl3) {Hidden layer 3};
    \node[annot,above of=H3-1, node distance=0.35cm] (w3b3) {};
        \node[annot,left of=w3b3, node distance=0.75cm] (w3b3t) {$\mb{W}_3,\mb{b}_3$};
    \node[annot,left of=hl1] {Input layer};
    \node[annot,right of=hl3] {Output layer};
        \node[annot,right of=w3b3, node distance=0.75cm] (w4b4t) {$\mb{W}_4,\mb{b}_4$};
\end{tikzpicture}
}
    \caption{Classification neural network with input, hidden and output layers. The input layer takes the vector $\mb{u}$ as input. A weight matrix $\mb{W}$ and bias $\mb{b}$ is applied between each layer. At the neurons of each hidden layer the non-linear ReLU activation function is applied. For the binary classification the magnitude of the two outputs are compared, i.e. $z_1 \geq z_2$ or $z_1 < z_2$. }
    \label{NN_structure}
\end{figure}
The general architecture of the classification neural network is illustrated in Fig.~\ref{NN_structure}. The neural network is defined by a number $K$ of fully-connected hidden layers, that each consist of $N_k$ number of neurons with $k = 1,...,K$. The neural network input vector $\mb{u}$ has dimension $N_0 \times 1$ and the output vector $\mb{z}$ has dimension $N_{K+1} \times 1$. Here, the dimension of $\mb{z}$ is $2 \times 1$, as we consider binary classification. The input to each layer $\mb{\hat{y}_k}$ is defined as:
\begin{align}
    \mb{\hat{y}}_{k+1} = \mb{W}_{k+1} \mb{y}_k + \mb{b}_{k+1} & \quad \forall k  = 0,1, ..., K-1 \label{NN_1}
\end{align}
where $\mb{y_0} = \mb{u}$ is the input to the neural network. The weight matrices $\mb{W}_{k}$ have dimensions $N_{k+1} \times N_k$ and the bias vector $\mb{b}$ has dimension $N_{k+1} \times 1$. Each neuron in the hidden layer applies a non-linear activation function to the input. Here, we assume the ReLU activation function: 
\begin{align}
    \mb{y}_k = \max (\mb{\hat{y}}_{k},0) & \quad \forall k  = 1, ..., K \label{NN_2}
\end{align}
The ReLU activation function in \eqref{NN_2} outputs $0$ if the input $\mb{\hat{y}}$ is negative, otherwise it propagates the input $\mb{\hat{y}}$. Note that the $\max$ operator is applied element-wise to the vector $\mb{\hat{y}_{k}}$. The majority of recent neural network applications uses the ReLU function as activation function as it has been found to accelerate neural network training \cite{glorot2011deep}. The main challenge of implementing other activation functions is their mixed-integer reformulation. While it is straightforward for (15) of ReLU, for other activation functions a direct reformulation might not be possible. However, one could approximate other activation functions (e.g. SiLU) with piecewise linear functions. The output of the neural network is:
\begin{align}
    \mb{z} = \mb{W}_{K+1} \mb{y}_{K} + \mb{b}_{K+1} \label{NN_3}
\end{align}
For binary classification, based on a comparison of the magnitude of the neural network output, we can either classify the input as belonging to the first class $z_1 \geq z_2$ or the second class $z_2 > z_1$. Here, the first class  $z_1 \geq z_2$ corresponds to the prediction that the input is in the feasible space $\mb{u} \in \mathcal{F}$, and the second class $z_2 > z_1$ to the prediction that the input is not in the feasible space $\mb{u} \notin \mathcal{F}$.

To train the neural network a dataset of labeled samples is required. The performance of the neural network highly depends on the quality of the dataset used.To encode the feasible space $\mathcal{F}$ of a general optimization problem \eqref{obj1}--\eqref{intract}, we sample $\mb{u}$ from the set $\mathcal{U}$ and determine whether the sample is feasible or not. For the application to the \mbox{N-1} security and small-signal stability constrained AC-OPF in \eqref{obj}--\eqref{SS2} this requires to sample $\mb{u} = [\mb{p_g^0}\,\mb{v_g^0}]^T$ from within the limits defined in \eqref{con_p}--\eqref{con_v}, and then test feasibility with respect to all constraints \eqref{con_p}--\eqref{SS2}.  If available, historical data of secure operating points can be used by the transmission system operator in conjunction with dataset creation methods. To achieve satisfactory neural network performance, the goal of the dataset creation is to create a balanced dataset of feasible and infeasible operating points which at the same time describe the feasibility boundary in detail. This is particularly important in AC-OPF applications as large parts of the possible sampling space lead to infeasible operating points \cite{thams2019efficient}. While the dataset creation is not the scope of this work, several approaches \cite{thams2019efficient, venzke2019efficient} focus on efficient methods to create datasets.

Before training of the neural network, the dataset is split into a training and test data set. During training of classification networks, the cross-entropy loss function is minimized using stochastic gradient descent \cite{tensorflow2015-whitepaper}. This penalizes the deviation between the predicted and true label of the training dataset. The weight matrices $\mb{W}$ and biases $\mb{b}$ are updated at each iteration of the training to minimize the loss function. After training, the generalization capability of the neural network is evaluated by calculating the accuracy and other relevant metrics on the unseen test data set.

\subsection{Exact Mixed-Integer Reformulation of Trained Neural Network}
With the goal of using the trained neural network in an optimization framework, following the work in  \cite{tjeng2017evaluating}, we first need to reformulate the maximum operator in \eqref{NN_2} using binary variables $\mb{b_k} \in {0,1}^{N_k}$ for all $k = 1, ...,K$:
\begin{subnumcases}{\mb{y}_k = \max(\hat{\mb{y}}_k,0)\Rightarrow}
\mb{y}_k  \leq \hat{\mb{y}}_k - \hat{\mb{y}}^{\text{min}}_k  (1-\mb{b}_k) \label{ReLU1} \\ 
\mb{y}_k  \geq \hat{\mb{y}}_k \label{ReLU2}   \\
\mb{y}_k  \leq \hat{\mb{y}}^{\text{max}}_k \mb{b}_k  \label{ReLU3}  \\
\mb{y}_k   \geq \mb{0}  \label{ReLU4}  \\
\mb{b}_k \in \{0,1\}^{N_k} \label{ReLUe}
\end{subnumcases}
We introduce one binary variable for each neuron in the hidden layers of the neural network. In case the input to the neuron is $\mb{\hat{y}} \leq 0$ then the corresponding binary variable is 0 and \eqref{ReLU3} and \eqref{ReLU4} constrain the neuron output $\mb{y}_k$ to 0. Conversely, if the input to the neuron is $\mb{\hat{y}} \geq 0$, then the binary variable is 1 and \eqref{ReLU1} and \eqref{ReLU2} constrain the neuron output $\mb{y}_k$ to the input $\hat{\mb{y}}_k$. Note that the minimum and maximum bounds on the neuron output $\hat{\mb{y}}^{\text{min}}$ and $\hat{\mb{y}}^{\text{max}}$ have to be chosen large enough to not be binding \emph{and} as small as possible to facilitate tight bounds for the mixed-integer solver. We compute suitable bounds using interval arithmetic \cite{tjeng2017evaluating}. To maintain scalability of the resulting MILP, several approaches have been proposed in literature including weight and ReLU pruning \cite{xiao2018training}. Here, we follow \cite{xiao2018training} and prune weight matrices $\mb{W}$ during training, i.e. we gradually enforce a defined share of entries to be zero.

\subsection{Mixed-Integer Non-Linear Approximation}
Based on the exact mixed-integer reformulation of the trained neural network with \eqref{NN_1}, \eqref{NN_3} and \eqref{ReLU1}--\eqref{ReLUe}, we can approximate solutions to the intractable problem \eqref{obj}--\eqref{SS2} by solving the following tractable mixed-integer non-linear optimization problem instead:
\begin{alignat}{2}
\min_{\begin{smallmatrix} \mb{p}^0,\mb{q}^0,\mb{v}^0,\bm{\theta}^0 \\
\hat{\mb{y}},\mb{y},\mb{z}\end{smallmatrix}} \, \, & f(\mb{p_g^0}) && \label{obj_MINLP}\\
    \text{s.t.}  \quad& \mb{p_g^{\text{min}}} \leq \mb{p_g^0} \leq \mb{p_g^{\text{max}}} &&  \label{con_p_MINLP} \\
    & \mb{v_g^{\text{min}}} \leq \mb{v_g^0} \leq \mb{v_g^{\text{max}}} &&  \label{con_v_MINLP} \\
    & \mb{s}_{\text{balance}} (\mb{p}^0,\mb{q}^0,\mb{v}^0,\bm{\theta}^0) = \mb{0} &&  \label{con_acpf_MINLP} \\
          & \mb{\hat{y}_{1}} = \mb{W}_{1} [\mb{p_g^0}\,\mb{v_g^0}]^T  + \mb{b_{1}} && \label{NN_re1} \\
       & \mb{\hat{y}_{k}} = \mb{W}_{k} \mb{y_{k-1}} + \mb{b_{k}} && \, \forall k  = 2, ..., K \\
       & \eqref{ReLU1}-\eqref{ReLUe} && \,\forall k  = 1, ..., K \\
&     \mb{z} = \mb{W}_{K+1} \mb{y}_{K} + \mb{b}_{K+1}  && \label{NN_ree} \\
& z_1 \geq z_2 \label{class} && 
\end{alignat}
The inequality constraints \eqref{con_p_MINLP}--\eqref{con_v_MINLP} provide upper and lower bounds on the control variables $\mb{u} = [\mb{p_g^0}\,\mb{v_g^0}]^T$. For the intact system state, \eqref{con_acpf_MINLP} ensures the non-linear AC power balance for each bus. The exact mixed-integer reformulation of the neural network is given in \eqref{NN_re1}--\eqref{NN_ree}. The constraint on the neural network output $\mb{z}$ in \eqref{class} ensures that the neural network predicts that the input $[\mb{p_g^0}\,\mb{v_g^0}]$ belongs to the feasible space $\mathcal{F}$ of the original problem in \eqref{obj}--\eqref{SS2}. Note that the neural network \eqref{NN_re1}--\eqref{class} encodes all constraints related to \mbox{N-1} security and small-signal stability, and eliminates all related optimization variables. The remaining non-linear constraint in \eqref{con_acpf_MINLP} enforces the non-linear AC power flow balance for the intact system state and requires to maintain the state variables for the intact system state only. While the number of optimization variables in \eqref{obj_MINLP}--\eqref{class} has been substantially reduced compared to \eqref{obj}--\eqref{SS2}, the resulting optimization problem is a mixed-integer non-linear problem, which are in general hard to solve. In the following, we will propose a systematic procedure to handle the non-linear equality constraint in \eqref{con_acpf_MINLP}.

\subsection{Mixed-Integer Linear Approximation}\label{sec:MILP}
The non-linear AC power flow equations described by the nodal power balance in \eqref{con_acpf_MINLP} can be summed over all buses. Then, we take the real part and write the summed active power balance as:
\begin{align}
    \sum_{\mathcal{G}} \mb{p_g^0} +  \sum_{\mathcal{N}} \mb{p_d^0} + p_{\text{losses}} (\mb{p}^0,\mb{q}^0,\mb{v}^0,\bm{\theta}^0) = 0 \label{nonlinearACPF1}
\end{align}
The first term $\sum_{\mathcal{G}} \mb{p_g^0}$ is the sum of active power generation and the second term $\sum_{\mathcal{N}} \mb{p_d^0}$ the sum of the active power loading of the system where the parameter $\mb{p_d^0}$ represent the active power load at each bus in $\mathcal{N}$. The third term $p_{\text{losses}}$ encapsulates the non-linearity and represents the active power losses. We propose to use an iterative linear approximation of the third non-linear term with a first-order Taylor expansion. At iteration $i+1$, we approximate \eqref{nonlinearACPF1} as:
\begin{align}
        \sum_{\mathcal{G}} \mb{p_g^0} +  \sum_{\mathcal{N}} \mb{p_d^0} + p_{\text{losses}}|_{i} + \nonumber \quad \quad \\
        \tfrac{\delta p_{\text{losses}}}{\delta \mb{p_g^0}}|_{i} (\mb{p_g^0} - \mb{p_g^0}|_{i}) + \tfrac{\delta p_{\text{losses}}}{\delta \mb{v_g^0}}|_{i} (\mb{v_g^0} - \mb{v_g^0}|_{i}) = 0 \label{nonlinearACPF}
\end{align}
The notation $|_{i}$ is shorthand for evaluated at the operating point of the $i$-th iteration, i.e. at $(\mb{p}^0,\mb{q}^0,\mb{v}^0,\bm{\theta}^0)|_{i}$. At the current operating point, we evaluate the value of the losses $p_{\text{losses}}|_{i}$ and the gradients with respect to the active generator dispatch $ \tfrac{\delta p_{\text{losses}}}{\delta p_g^0}|_{i}$ and the generator voltages $\tfrac{\delta p_{\text{losses}}}{\delta v_g^0}|_{i}$, respectively. Then, we approximate the value of the losses as a function of both the active generator dispatch and the generator voltages using the first-order Taylor expansion. 

The iterative scheme has the following steps. To initialize for iteration $i=0$, we linearize around a known operating point $(\mb{p}^0,\mb{q}^0,\mb{v}^0,\bm{\theta}^0)|_{i=0}$, e.g. the solution to the AC-OPF or \mbox{N-1} security-constrained AC-OPF problem. Then, we solve the following mixed-integer linear program for iteration $i$:
\begin{align}
    \min_{\mb{p_g^0}, \mb{v_g^0}, \hat{\mb{y}},\mb{y},\mb{z}} \quad & f(\mb{p_g^0}) \label{MILP1} \\
   \text{s.t.}  \quad & \eqref{con_p_MINLP},\, \eqref{con_v_MINLP},\, \eqref{NN_re1}-\eqref{class},\, \eqref{nonlinearACPF} \label{MILP2} 
\end{align}
 By iteratively linearizing the non-linear equation in \eqref{con_acpf_MINLP}, we solve MILPs instead of MINLPs. MILPs can be solved at a fraction of the time required for MINLPs as the convexity of the problem without integer variables allows for a significantly improved pruning of the branch-and-bound procedure. Based on the result of this optimization problem, we subsequently run an AC power flow for the intact system state to recover the full power system state and the losses. The iterative scheme converges if the change between the active power losses between iteration $i$ and $i-1$ is below a defined threshold $\rho$: $\tfrac{p_{\text{losses}}|_{i} - p_{\text{losses}}|_{i-1}}{p_{\text{losses}}|_{i}}  \leq \rho$. Otherwise, we resolve \eqref{MILP1}--\eqref{MILP2} with the loss parameters in \eqref{nonlinearACPF} updated. In Section~\ref{results_detailed}, we demonstrate fast convergence of this method.

\subsection{Ensuring Feasibility of Solutions}
\begin{figure}
    \centering
  \includegraphics[width=7.5cm]{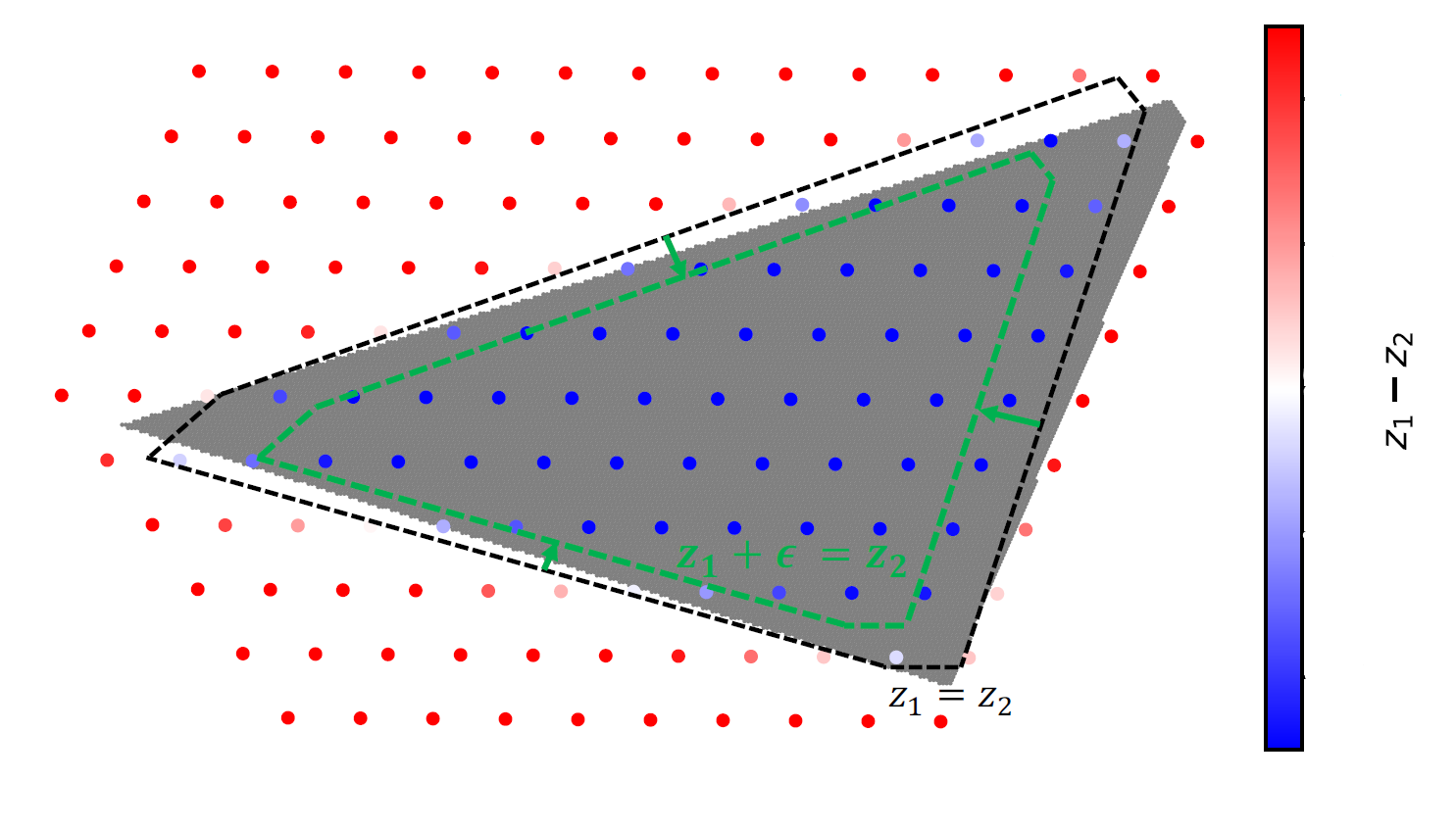}
    \caption{Illustrative reduction of predicted feasible space: Red dots are predicted infeasible, blue dots are predicted feasible. The grey area represents the true feasible space. The black dotted line represents the predicted feasibility boundary. The green dotted line is the predicted feasibility boundary after introducing a suitable $\epsilon$ in the classification decision $z_1 \geq z_2 + \epsilon$.}
    \label{Fig_Feas}
\end{figure}
The classification neural network is trained on a finite number of data samples and with a finite number of neurons. As a result, there might be a mismatch between the prediction of the feasibility boundary and the actual true feasibility boundary. This could lead to falsely identifying an operating point as feasible while it is in fact not feasible. Note that feasibility refers to satisfication of both static and dynamic security criteria. To control the conservativeness of the neural network prediction we introduce an additional constant factor in the classification and replace \eqref{class} in the MILP formulation \eqref{MILP1}--\eqref{MILP2} with: 
\begin{align}
    z_1 \geq z_2 + \epsilon \label{eps_control}
\end{align}
Adding the term $\epsilon$ to the right hand side in \eqref{eps_control} effectively shrinks the size of the predicted feasible space. This reduces the risk that operating points close to the feasibility boundary are falsely classified as feasible. But this can lead to an increase of the cost of operation, i.e. there is a trade-off between ensuring feasibility and optimality. In Fig.~\ref{Fig_Feas} we illustrate the reduction of the predicted feasible space by introducing $\epsilon$. In Section~\ref{results_detailed}, we quantify the trade-off between ensuring feasibility and obtaining a cost-optimal solution.
\section{Simulation and Results}
In the following, we demonstrate our proposed methodology using an IEEE 14-bus system and considering combined \mbox{N-1} security and small-signal stability as security criteria. 

\subsection{Simulation Setup}
As test case, we use the IEEE 14-bus system from \cite{zimmerman2010matpower} consisting of 5 generators, 11 loads and 20 lines with the following modifications: For the \mbox{N-1} security criterion, we consider all possible line outages except of the lines connecting buses 7 to 8 and 6 to 13, similar to previous works \cite{thams2019efficient}, as including these outages would render the problem infeasible. Furthermore, we enlarge the apparent branch flow limits in \eqref{con_s} by 30\% and the reactive power generator limits in \eqref{con_q} by 25\% to be able to obtain feasible solutions. To build the system matrix $\mb{A}$ of the small signal stability model in \eqref{SS1}, we use a tenth-order synchronous machine model and follow standard modeling procedures outlined in \cite{milano2010power}. We use Mathematica 
to derive the small signal model, MATPOWER AC power flows to initialize the system matrix \cite{zimmerman2010matpower}, and Matlab to compute its eigenvalues and damping ratio, and assess the small-signal stability for each operating point and contingency. Note that we require a minimum damping ratio $\gamma_{\text{min}}$ of 3\%. 

To create the dataset of labeled operating points, we discretize the set $\mathcal{U}$ within the minimum and maximum generator limits in \eqref{con_p_MINLP}. For each sample, we compute the feasibility w.r.t. combined \mbox{N-1} security and small-signal stability. As large parts of this set lead to infeasible solutions, we resample around identified feasible solutions. As a result, we obtain a dataset of 45'000 datapoints with 50.0\% feasible and 50.0\% infeasible samples. We split the dataset into 80\% training and 20\% test set and train a neural network with 3 hidden layers and 50 neurons in each hidden layer using TensorFlow \cite{tensorflow2015-whitepaper}. During training, we gradually increase the enforced sparsity of the weight matrices to 50\%, i.e. 50\% of the entries of $\mb{W}$ have to be zero. This increases the tractability of the resulting MILPs significantly \cite{xiao2018training}. Finally, we obtain a trained neural network with a predictive accuracy of 99\% on the test dataset, showing good generalization capability of the neural network.

We formulate the mixed-integer linear programs \eqref{MILP1}--\eqref{MILP2} in YALMIP \cite{lofberg2004yalmip} and solve it using Gurobi. We evaluate the performance of our methodology for 125 random objective functions. For these we draw random linear cost coefficients for each of the five generators between $5 \tfrac{\$}{\text{MW}}$ and $20 \tfrac{\$}{\text{MW}}$ using Latin hypercube sampling. We initialize the loss approximation in \eqref{nonlinearACPF} with the solution to the AC-OPF for each random cost function. The computational experiments are run on a laptop with an Intel Core i7 CPU @ 2.70 GHz, 8 GB RAM, and 64-bit operating system.

\subsection{Simulation Results}
\label{results_detailed}
\begin{table*}[]
    \caption{Comparison of problem formulation and type, generation cost, and share of feasible instances for 125 instances with random cost functions}
    \vspace{0.3cm}
    \centering
    \begin{tabular}{c c c c}
    \toprule 
       Problem  & Problem & Generation   & Share of feasible\\
        formulation & type & cost (\$) & instances (\%)\\
         \midrule 
        AC-OPF  & NLP & 2425.94  & 35.2 \\
          + \mbox{N-1} security  & NLP & 2565.13 & 35.2 \\
       + small-signal stability ($\epsilon = 0$) & MILP & 2942.83  &  52.8 \\ 
       + small-signal stability ($\epsilon = 7$) & MILP & 2949.78  &  84.0 \\ 
         + small-signal stability ($\epsilon = 8$) & MILP & 2950.76  & 100.0 \\       
         \bottomrule 
    \end{tabular}
    \label{Comp_Cost_Time}
\end{table*}

In Table~\ref{Comp_Cost_Time}, we compare the solutions to the conventional AC-OPF and \mbox{N-1} security constrained AC-OPF (both solved using MATPOWER and IPOPT \cite{wachter2006implementation}) with our proposed approach encoding the feasible space to MILPs in terms of problem formulation and type, generation cost, solver time and share of feasible instances for 125 instances with random cost functions. Note that random cost functions allow us to explore the feasible region better, as they result to different optimal points. At the same time, random cost functions do not affect the feasible space. Both the conventional AC-OPF and \mbox{N-1} security constrained AC-OPF do not return a feasible solution for 64.8\% of the 125 random test instances, requiring computationally expensive post-processing. Following our proposed approach, where we include the dynamic constraints in the optimization problem encoded as a MILP, for a value of $\epsilon =0$, we decrease this share to 47.2\%. Introducing a value of $\epsilon =8$ in the classification decision of the neural network in \eqref{eps_control} allows us to obtain feasible solutions for all test instances, i.e. all obtained solutions satisfy \mbox{N-1} security and small-signal stability criteria. Note that we select the value of $\epsilon =8$ based on a sensitivity analysis that we explain in detail in the following paragraphs. The iterative loss approximation in section \ref{sec:MILP} with $\rho = 1\%$ converges rapidly within 7 iterations on average. This shows good performance of the proposed approach to handle the equality constraint.

 In Table~\ref{Comp_Cost_Time}, the average cost increase of including \mbox{N-1} security evaluates to 5.7\%, and including small-signal stability leads to an additional increase of 15.6\% which is expected. Notably, the additional cost increase by introducing a value of $\epsilon = 8$, i.e. increasing the conservativeness of the neural network prediction, evaluates only to 0.033\%, while it increases the share of feasible instances from 84.0\% to 100\%. 
 Note that in terms of scalability, our proposed approach requires only to include the control variables of the intact system state, and depends mainly on the tractability of the MILP reformulation of the neural network. In contrast to that, the complexity of the conventional \mbox{N-1} security constrained AC-OPF increases exponentially for large scale instances requiring iterative solution schemes \cite{capitanescu2011state}. The scalability of the proposed approach is not affected by the power system size and number of contingencies, assuming a given fixed neural network size. 

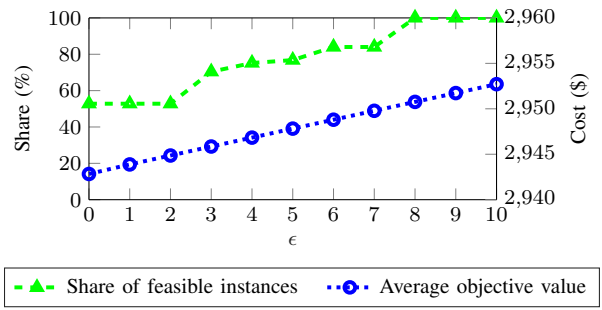
\begin{figure}
    \centering
    \footnotesize
%
%
\begin{tikzpicture}

\begin{axis}[%
width=7cm,
height=4cm,
xmin=0,
xmax=10,
xtick={ 0,  1,  2,  3,  4,  5,  6,  7,  8,  9, 10},
xlabel style={font=\color{white!15!black}},
xlabel={$\epsilon$},
ymin=0,
ymax=100,
ylabel={Share (\%)},
 axis y line*=left,
grid style={black},
legend style={at={(0.03,0.97)}, anchor=north west, legend cell align=left, align=left, draw=white!15!black}
]
\addplot [color=green, dashed, line width=1.5pt, mark=triangle, mark options={solid, green}]
  table[row sep=crcr]{%
0	52.8\\
1	52.8\\
2	52.8\\
3	70.4\\
4	75.2\\
5	76.8\\
6	84\\
7	84\\
8	100\\
9	100\\
10	100\\
};

\end{axis}
\begin{axis}[
width=7cm,
height=4cm,
xmin=0,
xmax=10,
xtick={ 0,  1,  2,  3,  4,  5,  6,  7,  8,  9, 10},
  axis y line*=right,
  axis x line=none,
  ymin=2940, ymax=2960,
  ylabel= Cost (\$),
]
\addplot [color=blue, dotted, line width=1.5pt, mark=o, mark options={solid, blue}]
  table[row sep=crcr]{%
0	2942.83\\
1	2943.88\\
2	2944.86\\
3	2945.85\\
4	2946.83\\
5	2947.82\\
6	2948.80\\
7	2949.78\\
8	2950.76\\
9	2951.73\\
10	2952.71\\
};

\end{axis}
\end{tikzpicture}%
    \begin{center}
			         \begin{tikzpicture}
    \begin{axis}[%
    hide axis,
    xmin=1,
    xmax=2,
    ymin=0,
    ymax=0.1,
    legend style={draw=white!15!black,legend cell align=left},
    legend columns=2
    ]
    \addlegendimage{color=green, dashed, line width=1.5pt, mark=triangle, mark options={solid, green}}
    \addlegendentry{Share of feasible instances \, \,};
        \addlegendimage{color=blue, dotted, line width=1.5pt, mark=o, mark options={solid, blue}}
    \addlegendentry{Average objective value};
    \end{axis}
    \end{tikzpicture}
     \end{center}
    \caption{For 125 random cost functions, we show the share of feasible instances and average cost function as function of parameter $\epsilon$.}
              \label{share_of_stable_points_vs_eps}
\end{figure}


To explore the trade-off between conservativeness of the neural network boundary prediction (through increasing $\epsilon$) and the resulting generation cost increase, Fig.~\ref{share_of_stable_points_vs_eps} shows the share of feasible instances and the average objective value for $\epsilon$ ranging from 0 to 10. The share of feasible instances increases in two jumps from $\epsilon = 3$ to $\epsilon = 4$ (from 52.8 \% to 70.4\%) and from $\epsilon = 7$ to $\epsilon = 8$, from 84.0\% to 100\%. As an alternative to the proposed heuristic of identifying $\epsilon$, our future work is directed towards robust retraining of neural networks \cite{venzke2019verification}.

Our proposed approach has the following limitations. First, the approach may incur optimality loss, due to the approximation errors. However, considering that the constraints about small-signal stability are intractable for any conventional optimization method, all methods are bound to use some sort of approximation. So far, no method exists in the literature that introduces approximations for dynamic security and guarantees no optimality loss. At the same time, a tractable, efficient and suboptimal solution can still provide useful information to the grid operator, compared to a solution that would not consider dynamic security at all. Second, if we apply this approach to different power grids, we will need to train different NNs, as, at the moment, it is difficult to train NNs that can generically apply to any power system. First approaches to train more general neural networks have already been proposed using e.g. Graph Neural Networks \cite{Marot_GNN}, \cite{Marot_GNN_EPSR}. Further developing such approaches will be the object of future work.



\section{Conclusion}
We introduce a framework that can efficiently capture \emph{previously intractable} optimization constraints and transform them to a mixed-integer linear program, through the use of neural networks. First, we encode the feasible space which is characterized by both tractable and intractable constraints, e.g. constraints based on differential equations, to a neural network. Leveraging an exact mixed-integer reformulation of the trained neural network, and an efficient method to include non-linear equality constraints, we solve mixed-integer linear programs that can efficiently approximate non-linear optimization programs with previously intractable constraints. We apply our methods to the AC optimal power flow problem with dynamic security constraints. For an IEEE 14-bus system, and considering a combination of \mbox{N-1} security and small-signal stability, we demonstrate how the proposed approach allows to obtain cost-optimal solutions which at the same time satisfy both static and dynamic security constraints. To the best of our knowledge, this is the first work that utilizes an exact transformation to convert the information encoded in neural networks to set of mixed integer-linear constraints and use it in an optimal power flow problem. Future work is directed towards utilizing efficient dataset creation methods \cite{venzke2019efficient}, increasing robustness of classification neural networks to boost the applicability of this approach \cite{venzke2019verification}, determining critical system indices using physics-informed neural networks \cite{georgios}, and learning worst-case guarantees for neural networks \cite{venzke2020,nellikkath}.

\bibliographystyle{IEEEtran}

\bibliography{Bib}

\end{document}